\documentclass{ayss_en}
\usepackage{graphicx}
\usepackage{amsmath,amsfonts,verbatim,graphicx,float}

\title{
  PROVING THE CEP WITH COMPACT STARS?\thanks{Work supported by CompStar, a 
research networking programme of the European Science foundation and by 
CompStar-POL a project of the Polish MNiSW supporting it as well as by NCN 
under grant number DEC-2011/02/A/ST2/00306 and by RFBR under grant number 
11-02-01538a.}
}
\author{
  D.E. Alvarez-Castillo\superscript{\small 1,}\superscript{\small 2}\thanks{alvarez@theor.jinr.ru} and
 D.~Blaschke\superscript{\small 1,}\superscript{\small 3,}\superscript{\small 4}
\thanks{blaschke@ift.uni.wroc.pl}\\
\superscript{\small 1}
Bogoliubov  Laboratory of Theoretical Physics,
JINR Dubna, 141980 Dubna, Russia\\
\superscript{\small 2}
Instituto de F\'{\i}sica,
Universidad Aut\'onoma de San Luis Potos\'{\i},
S.L.P. 78290, M\'exico\\
\superscript{\small 3}
Institute for Theoretical Physics,
University of Wroc{\l}aw,
50-204 Wroc{\l}aw, Poland\\
\superscript{\small 4}
	Fakult\"at f\"ur Physik,
	Universit\"at Bielefeld,
	33501 Bielefeld, Germany
}

\begin{document}
  \maketitle
  \begin{abstract}
   We present a model for hybrid compact stars composed of a quark core and a 
hadronic mantle with an abrupt first order phase transition at the interface 
and in accordance with the latest astrophysical measurements of two 2 M$_\odot$
pulsars. 
We demonstrate the possibility of a disconnected mass-radius sequence 
(third family) of high-mass pulsars as a distinct feature due to a large 
jump $\Delta \epsilon$ in the energy density of the first order phase 
transition setting in at $\epsilon_{\rm crit}\approx 500$ MeV/fm$^3$ and
fulfilling $\Delta \epsilon/\epsilon_{\rm crit} > 0.6$.
We conclude that the measurement of so called \textit{twin} compact stars 
%(same mass but different intermal structure) 
at high mass ($\sim$ 2 M$_\odot$) would support the existence of a first order 
phase transition in symmetric matter at zero temperature 
entailing the existence of a 
critical end point (CEP) in the QCD phase diagram.
  \end{abstract}
  
\section{Introduction}
Recently the study of neutron stars has become an ever more active field of 
research since it complements the investigations of matter under extreme 
conditions in heavy-ion collisions and in ab-initio lattice QCD simulations. 
The interiors of compact stars are unique places to investigate the otherwise
inaccessible cold, dense equation of state (EoS) which determines the shape of 
of stable compact star sequence(s) in the mass - radius diagram as well as 
other properties like the composition, the moment of inertia etc. 
In this regard, recently the masses of two heavy compact stars have been 
precisely measured: 1.97 $\pm $ 0.04 M$_{\odot}$ for PSR J1614-2230 
\cite{Demorest:2010bx} and 2.01 $\pm $ 0.04 M$_{\odot}$ for  PSR J0348+0432 
\cite{Antoniadis:2013pzd}, 
presenting a stringent constraint on models for the high density EoS. 
Would it be possible to gain evidence for a strong first order 
deconfinement transition in compact stars this would prove the existence 
of at least one CEP in the QCD phase diagram since in the high temperature 
region explored by lattice QCD simulations  \cite{Kaczmarek:2011zz}
the transition is crossover.

\section{Massive hybrid stars \& twins}

\begin{figure}[!th]
\center
\includegraphics[width=8cm,height=8cm]{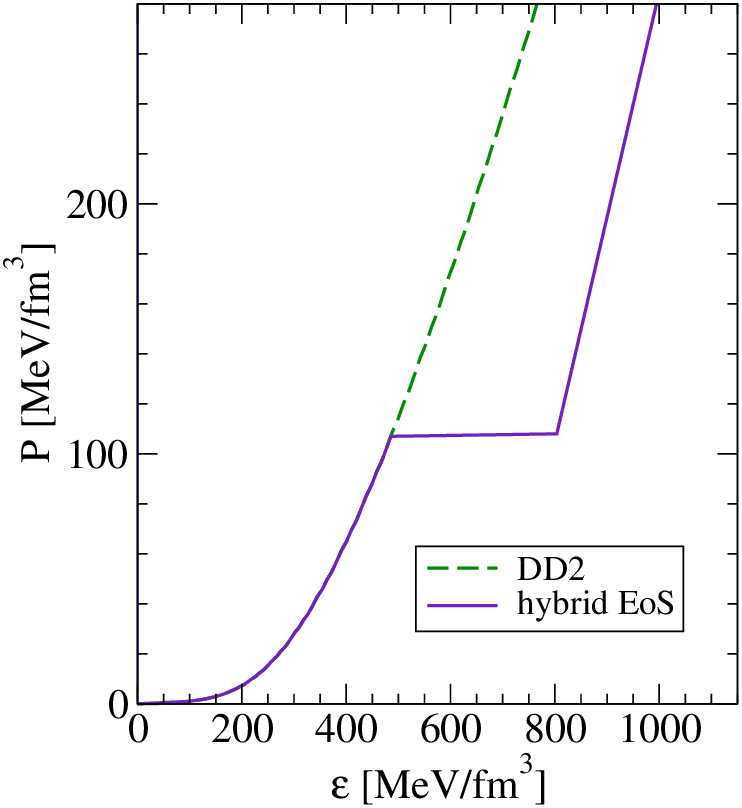}
\includegraphics[width=8cm,height=8cm]{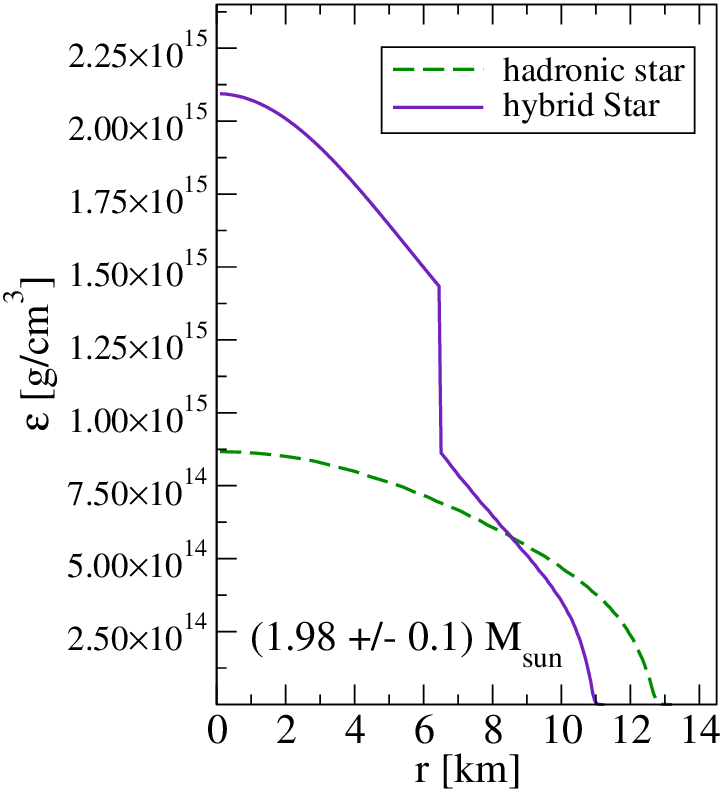}
\caption{\label{EoS}
Left panel: Hadronic EoS (DD2, green dashed line) vs. hybrid one 
(Eq.~\ref{eos}, indigo solid line).
Right panel:
Energy density profile of 2 M$_\odot$ twins: a hadronic (green dashed line) 
and a hybrid (indigo solid line) compact star. 
%Clearly the latter presents an abrupt density jump characteristic of the 
%first order phase transition.
 }
\end{figure}

\begin{figure}[!th]
\center
\includegraphics[width=12cm]{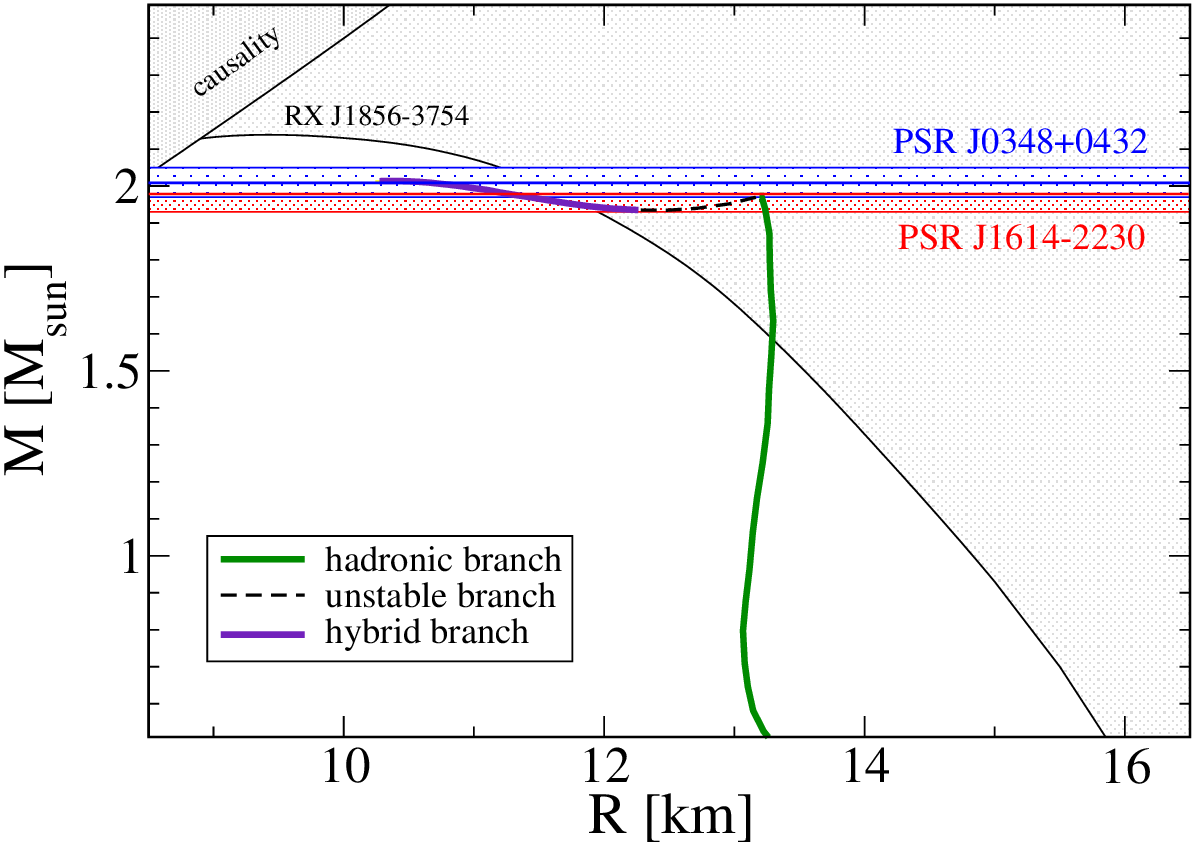}
\caption{\label{MvsR}
Mass-radius diagram for hybrid star EoS (\ref{eos}). 
The almost vertical branch of pure hadronic stars (green solid line) 
is separated from the almost horizontal hybrid star branch (indigo solid line) 
by an unstable branch (black dashed line).
Recently measured masses for two 2 M$_\odot$ pulsars shown by red and blue 
hatched regions.
Also shown: causality constraint (upper left forbidden area) and 
the minimal radius constraint from RXJ 1856-3754 \cite{Trumper:2003we}.}
\end{figure}

A first order phase transition in neutron star matter can take place just as 
in symmetric matter where it is searched for in heavy ion collisions. 
Adopting the setting of \cite{Alford:2013aca,Zdunik:2012dj}, we construct 
hybrid stars with a hybrid EoS composed of a given hadronic EoS, here 
DD2~\cite{Typel:1999yq}, and a quark matter EoS parametrized by its 
squared speed of sound $c_{\rm QM}^2$ which pretty well describes 
\cite{Zdunik:2012dj} results of a color superconducting NJL 
model~\cite{Lastowiecki:2011hh}
\begin{equation}
\label{eos}
P(\varepsilon)= 
P_{\rm DD2}(\varepsilon)\Theta(\varepsilon_{\rm crit}-\varepsilon)
+c_{\rm QM}^2~\varepsilon~\Theta(\varepsilon-\varepsilon_{\rm crit}
-\Delta\varepsilon)~.
\end{equation}
The critical energy density $\varepsilon_{\rm crit}$ and the discontinuity 
$\Delta\varepsilon$ complete the EoS model which is capable of describing
compact star sequences with a third family of stars in the 
mass-radius diagram
\cite{Gerlach:1968zz,Kampfer:1981yr,Schertler:2000xq,Glendenning:2000gh}.
Searching for sequences obeying the mass constraint 
\cite{Demorest:2010bx,Antoniadis:2013pzd} we obtain a
quasi-horizontal branch disconnected from the almost vertical neutron 
star branch, as a consequence of a strong phase transition. 
Fig.~\ref{EoS} shows the EoS (\ref{eos}) for the parameters: 
$c_{\rm QM}^{2}= 0.94$, $\Delta\varepsilon = 0.67~\varepsilon_{\rm crit}$ and 
$\varepsilon_{\rm crit}=485$ MeV/fm$^3$. 
The latter corresponds to $P(\varepsilon_{\rm crit})=100$ MeV/fm$^3$ and a 
baryon density at the quark matter onset of 
$n_{\rm crit}= 2.9~n_0$ with $n_0=0.16$ fm$^{-3}$. 
Fig.~\ref{MvsR} shows the mass-radius relation for this hybrid EoS and the energy density profile for the example of a twin pair of 2 M$_{\odot}$ 
stars is given in Fig.~\ref{EoS} (right). 


\section{Conclusions}

In this work we have demonstrated for a simple hybrid star EoS model 
that the characteristic mass-radius relationship of nonrotating 
compact star configurations obtained by solving the TOV equations has
a third family of stable hybrid stars at high masses that is in accordance 
with astrophysical observations.
This includes in particular the constraint for the minimal radius 
from RXJ 1856-3754 and the new pulsar mass measurements for 
PSR J1614-2230 and PSR J0348+0432 at about 2 M$_\odot$.
Our result shows that stars of this high mass which are the best candidates 
for containing exotic matter (like quark matter) in their interiors
could actually be approximate mass twins: although having about the same mass
their interiors may show a vastly different composition:
while the lower density twin has a larger radius ($\sim 13$ km) and is made 
of nuclear matter only, the high-density twin is more compact 
($\sim 10.5 - 12$ km) with a phase transition at about half the stars radius.
The precise determination of compact star radii would allow to prove
or disprove whether the sequences of stable stars in the mass-radius diagram 
form a pattern like the one discussed here: with an almost vertical branch of 
hadronic stars and an almost horizontal branch of hybrid stars, possibly
disconnected from the former thus representing a ``third family'' of compact 
stars and allowing for high-mass star twins.
Identifying this pattern 
%the high-mass twin feature  
by observation 
%of a disconnected almost horizontal ``third family'' branch in the 
%mass-radius diagram 
would indicate that %low-temperature 
compact star matter 
undergoes a phase transformation with a large jump in energy density that could
correspond to a strong first order transition  in symmetric matter. 
Given the knowledge from lattice QCD that at zero baryon density
the QCD phase transition proceeds as a crossover, twins would then support the 
existence of a CEP in the QCD phase diagram, see also \cite{Blaschke:2013rma} 
for a microphysical study.

\section{Acknowledgements}
We gratefully acknowledge numerous discussions and collaboration work on the
topic addressed in this contribution with our colleagues, in particular with 
S. Benic, G. Contrera, H. Grigorian, O. Kaczmarek, T. Kl\"ahn, E. Laermann, 
R. {\L}astowiecki, M.C. Miller, G. Poghosyan, S.B. Popov, J. Tr\"umper, 
D.N. Voskresensky and F. Weber.


\begin{thebibliography}{9}

\bibitem{Demorest:2010bx} 
  P.~Demorest {\it et al.}, 
  %T.~Pennucci, S.~Ransom, M.~Roberts and J.~Hessels,
  %``Shapiro Delay Measurement of A Two Solar Mass Neutron Star,''
  Nature {\bf 467}, 1081 (2010).
  %[arXiv:1010.5788 [astro-ph.HE]].
  %%CITATION = ARXIV:1010.5788;%%

%\cite{Antoniadis:2013pzd}
\bibitem{Antoniadis:2013pzd} 
  J.~Antoniadis {\it et al.}, 
%P.~C.~C.~Freire, N.~Wex, T.~M.~Tauris, R.~S.~Lynch, M.~H.~van Kerkwijk, 
%M.~Kramer and C.~Bassa {\it et al.},
  %``A Massive Pulsar in a Compact Relativistic Binary,''
  Science {\bf 340}, 6131 (2013).
%  [arXiv:1304.6875 [astro-ph.HE]].
  %%CITATION = ARXIV:1304.6875;%%
    
%%%%%%%% lattice QCD crossover

%\cite{Kaczmarek:2011zz}
\bibitem{Kaczmarek:2011zz} 
  O.~Kaczmarek {\it et al.}, 
  %F.~Karsch, E.~Laermann, C.~Miao, S.~Mukherjee, P.~Petreczky, 
  %C.~Schmidt and W.~Soeldner {\it et al.},
  %``Phase boundary for the chiral transition in (2+1) -flavor QCD at small 
  %values of the chemical potential,''
  Phys.\ Rev.\ D {\bf 83}, 014504 (2011).
%  [arXiv:1011.3130 [hep-lat]].
  %%CITATION = ARXIV:1011.3130;%%

%%%%%%%%%% twins and first order EoS

%\cite{Alford:2013aca}
\bibitem{Alford:2013aca} 
  M.~G.~Alford, S.~Han and M.~Prakash,
  %``Generic conditions for stable hybrid stars,''
  arXiv:1302.4732 [astro-ph.SR].
  %%CITATION = ARXIV:1302.4732;%%  


%%DD2 citation:

%\cite{Typel:1999yq}
\bibitem{Typel:1999yq} 
  S.~Typel and H.~H.~Wolter,
  %``Relativistic mean field calculations with density dependent meson 
  %nucleon coupling,''
  Nucl.\ Phys.\ A {\bf 656}, 331 (1999).
  %%CITATION = NUPHA,A656,331;%%
  %175 citations counted in INSPIRE as of 28 Apr 2013

%\cite{Zdunik:2012dj}
\bibitem{Zdunik:2012dj} 
  J.~L.~Zdunik and P.~Haensel,
  %``Maximum mass of neutron stars and strange neutron-star cores,''
  arXiv:1211.1231 [astro-ph.SR].
  %%CITATION = ARXIV:1211.1231;%%
  %3 citations counted in INSPIRE as of 28 Apr 2013

%\cite{Lastowiecki:2011hh}
\bibitem{Lastowiecki:2011hh} 
  R.~Lastowiecki {\it et al.}, 
   %D.~Blaschke, H.~Grigorian and S.~Typel,
  %``Strangeness in the cores of neutron stars,''
  Acta Phys.\ Polon.\ Supp.\  {\bf 5}, 535 (2012)
  [arXiv:1112.6430 [nucl-th]].
  %%CITATION = ARXIV:1112.6430;%%


%%%Twins

%\cite{Gerlach:1968zz}
\bibitem{Gerlach:1968zz} 
  U.~H.~Gerlach,
  %``Equation of State at Supranuclear Densities and the Existence of a Third Family of Superdense Stars,''
  Phys.\ Rev.\  {\bf 172}, 1325 (1968).
  %%CITATION = PHRVA,172,1325;%%
  %35 citations counted in INSPIRE as of 28 Apr 2013aca

%\cite{Kampfer:1981yr}
\bibitem{Kampfer:1981yr} 
  B.~K\"ampfer,
  %``On The Possibility Of Stable Quark And Pion Condensed Stars,''
  J.\ Phys.\ A {\bf 14}, L471 (1981).
  %19 citations counted in INSPIRE as of 28 Apr 2013


%\cite{Schertler:2000xq}
\bibitem{Schertler:2000xq} 
  K.~Schertler {\it et al.}, 
  %C.~Greiner, J.~Schaffner-Bielich and M.~H.~Thoma,
  %``Quark phases in neutron stars and a 'third family' of compact stars as a 
  %signature for phase transitions,''
  Nucl.\ Phys.\ A {\bf 677}, 463 (2000).
%  [astro-ph/0001467].
  %%CITATION = ASTRO-PH/0001467;%%
  %77 citations counted in INSPIRE as of 28 Apr 2013

\bibitem{Glendenning:2000gh}
  N.~K.~Glendenning and C. Kettner, Astron.\ Astrophys.\ {\bf 353}, 795 (2000).
%%%%%%%%%%%%%%%%%%%%%%%5

%\cite{Trumper:2003we}
\bibitem{Trumper:2003we} 
  J.~E.~Tr\"umper {\it et al.}, 
  %V.~Burwitz, F.~Haberl and V.~E.~Zavlin,
  %``The Puzzles of RX J1856.5-3754: Neutron star or quark star?,''
  Nucl.\ Phys.\ Proc.\ Suppl.\  {\bf 132}, 560 (2004).
%  [astro-ph/0312600].
  %%CITATION = ASTRO-PH/0312600;%%

%\cite{Blaschke:2013rma}
\bibitem{Blaschke:2013rma} 
  D.~Blaschke {\it et al.}, 
  %D.~E.~A.~Castillo, S.~Benic, G.~Contrera and R.~Lastowiecki,
  %``Nonlocal PNJL models and heavy hybrid stars,''
  arXiv:1302.6275 [hep-ph].
  %%CITATION = ARXIV:1302.6275;%%


  \end{thebibliography}
\end{document}